\documentclass[aps,prx,onecolumn,superscriptaddress,10pt]{revtex4-2}
\usepackage[utf8]{inputenc}
\usepackage{amsmath}
\usepackage{hyperref}
\usepackage{mathdots}
\usepackage{dsfont}
\usepackage{amsfonts}
\usepackage{amssymb}
\usepackage{graphicx}

\usepackage{tikz}
\usetikzlibrary{quantikz}
\usepackage{multirow}
\usepackage{xcolor}
\usepackage{bbold}
\DeclareUnicodeCharacter{2009}{XXXX here is the little bugger XXXX}

\begin{document}

\title{Lattice Boltzmann-Carleman quantum algorithm and circuit for fluid flows at moderate Reynolds number}

\author{Claudio Sanavio}
\email{claudio.sanavio@iit.it}
\affiliation{Fondazione Istituto Italiano di Tecnologia\\
Center for Life Nano-Neuroscience at la Sapienza\\
Viale Regina Elena 291, 00161 Roma, Italy}

\author{Sauro Succi}
\affiliation{Fondazione Istituto Italiano di Tecnologia\\
Center for Life Nano-Neuroscience at la Sapienza\\
Viale Regina Elena 291, 00161 Roma, Italy}

\begin{abstract}
We present a quantum computing algorithm for fluid flows
based on the Carleman-linearization of the Lattice Boltzmann (LB) method.
First, we demonstrate the convergence of the classical Carleman procedure 
at moderate Reynolds numbers, namely for Kolmogorov-like flows. 
Then we proceed to formulate the corresponding quantum algorithm, including  the quantum
circuit layout and analyze its computational viability.
We show that, at least for moderate Reynolds numbers 
between $10$ and $100$, the Carleman-LB procedure can be successfully truncated 
at second order, which is a very encouraging result. 
We also show that the quantum circuit implementing the single time-step collision operator 
has a fixed depth, regardless of the number of lattice sites.
However,  such depth is of the order of ten thousands
quantum gates,  meaning that quantum advantage over classical computing 
is not attainable today,  but could be achieved in the near-mid term future.
The same goal for the multi-step version remains however an open topic
for future research.



\end{abstract}

\maketitle

\section{Introduction}\label{sec:I}

Quantum computing~\cite{nielsen_quantum_2010} holds promise to provide dramatic 
speed up to the solution of a number of major scientific  problems, including
advanced industrial and societal applications~
\cite{harrow_quantum_2009,albash_adiabatic_2018,giannakis_embedding_2022,mangini_quantum_2022}. 
The so-called quantum advantage stems from the deepest (and most counterintuitive)  
features of quantum mechanics, in particular, superposition and entanglement of quantum states 
which offer, at least in principle, the chance to exploit the full Hilbert space, scaling exponentially 
with the number of qubits, the smallest bit of quantum information. 
This feature provides a natural way out to the infamous  "curse of dimensionality", 
plaguing the simulation of most quantum many-body problems, both classical and quantum~\cite{georgescu_quantum_2014,tacchino_quantum_2020}.

Yet, realizing such a mind-boggling potential faces with a number of steep challenges, both conceptual
and technological, primarily fast decoherence and, even more so, the quantum noise affecting the 
operation of real-life quantum computers. 
Understandably, quantum computing to date has been directed mostly to quantum physics problems, 
featuring a one to one mapping between the physical system to be simulated and the quantum hardware
\cite{feynman_simulating_1982}.
Yet, there is a mounting interest in learning whether the potential of quantum computing 
can be put at use also for solving the most compelling problems in classical physics, as typically 
described by strongly nonlinear partial differential equations \cite{steijl_quantum_2019,todorova_quantum_2020,gaitan_finding_2020,zhao_quantum_2022,li_potential_2023}. 
In this respect, fluid turbulence stands out as a prominent candidate, both
in terms of fundamental physics and also in view of its pervasive applications 
in both natural and industrial phenomena.

This work inscribes precisely within the aforementioned scenario; we shall present a 
quantum algorithm, and the associated circuit, solving the basic (Navier-Stokes) equations 
which govern the physics of dissipative fluids. For reasons to be apparent in the sequel,
our strategy is  based on the Lattice Boltzmann formalism for fluid flows.

The paper is organized as follows. 
In Section~\ref{sec:II} we introduce the classical equation of motions. In Section~\ref{sec:III} we review the lattice Boltzmann method. In Section~\ref{sec:IV} we focus on 
the Carleman linearization. We introduce two ways to deal with the infinite number of variables, and 
we show how to conveniently cut down the size of the system of equations. 
We then show in Section~\ref{sec:V} an explicit analysis of the CL method and its 
performance on a classical computer. 
In Section~\ref{sec:VI} we define the embedding of the Carleman variables into the 
space of qubits and we explicitly construct the quantum circuit. 
This consists of two separate steps, to be applied in series,
the collision and the multi-streaming operator. 
Finally, in Section~\ref{sec:VII} we draw preliminary conclusions and draw a prospective outlook for
future works in this area.

\section{The equation of motion of classical fluids}\label{sec:II}

The Navier-Stokes equations (NSE), read as follows:

\begin{eqnarray}
\partial_t \rho + \partial_a(\rho u_a)&=&0\label{eq:continuity_equation}\\
\partial_t (\rho u_a) + \partial_b (\rho u_a u_b)&=&-\partial_a p
+ \partial_b \sigma_{ab}+F_a\label{eq:velocity_equation}
\end{eqnarray}

\noindent with $u_a$ being the macroscopic velocity of the fluid, $\rho$ the fluid density, $F_a$ the external force, $p$ the pressure of the fluid and $\sigma_{ab}$ the dissipative tensor. 
The latin indices $a,b$ run over the cartesian coordinates $x,y$ and $z$. 
The first line~\eqref{eq:continuity_equation} is the continuity equation, whereas the second line~\eqref{eq:velocity_equation} is a vectorial representation of the evolution of the macroscopic velocities. We use here the Einstein convention, by which repeated indices are summed upon. Eqs.~\eqref{eq:velocity_equation} are non-linear partial differential equations, and the strength  of the non-linearity bears heavily on our ability to solve the NSE, either analytically, or even using the most
advanced computational fluid dynamics (CFD) methods \cite{wendt_computational_2008}. 

The Reynolds number $\text{Re}$ is a measure of the non-linearity of the system and 
it is defined as the ratio between the inertial and viscous forces, $\text{Re}= u\cdot\nabla u/\nu\nabla^2 u$,
as given by the ratio: 
\begin{equation}
\text{Re} = \frac{|u|L}{\nu},
\end{equation}

\noindent where $|u|$ is the magnitude of the macroscopic velocity and $L$ is the global system size.
To be noted that the Reynolds number takes on very large numbers also under very mundane conditions, 
an ordinary car moving at a standard speed already features $Re \sim 10^7$, ten millions.
Given that the computational complexity of fluid turbulence scales like $Re^3$, this means that
$10^{21}$ active degrees of freedom need to be tracked in order to simulate the dynamics of a full car.
With order thousands floating point operation per degree of freedom, this yields $10^{24}$ floating
point operations, implying $10^6$ CPU/GPU seconds, about two weeks, to complete the simulation 
on an ideal Exascale computer. So much for an ordinary car.
Consider now the problem of numerical weather forecast, which implies Reynolds numbers easily in the
order of $10^{10}$, leading to an intractable problem for any foreseeable classical computer.
These simple figures speak clearly for the motivation to investigate the possibility of exploiting quantum
computers for simulating classical turbulence~\cite{succi_quantum_2023}.

CFD is a traditional forefront of computational science, with a ceaseless quest for 
better and more efficient computational methods.
In the last three decades, the Lattice Boltzmann method (LBM) , has gained a prominent role
in the CFD arena \cite{succi_lattice_1991,benzi_lattice_1992,succi_lattice_1993,
kruger_lattice_2017,falcucci_extreme_2021,succi_lattice_2022}.
In a nustshell, LBM is a stylized version of the Boltzmann equation which retains the essential
physics of fluids within a very efficient computational kinetic-theory harness. 
 
 It consists of two basic processes: streaming, by which particles move freely from one lattice site to the
 next, and collisions, whereby particles exchange mass, momentum and energy, so as to sustain the collective
 dynamics telling fluids apart from a "wild bunch" of independent particles.
 The streaming is minimally non local, as it connects single-cell neighbors, but linear, in fact
 {\it exact}, as no information is lost in moving information from one lattice site to another.
 Conversely, collisions hold the (quadratic) nonlinearity of the physics of fluids, but in a local   
 form, because, consistently with Boltzmann's kinetic theory, only particles in the same lattice site interact with each other.
This clearcut separation between nonlinearity and nonlocality is possibly the most profound hallmark of the
LBM, and the basic reason for its computational success especially on parallel computers.
It is therefore reasonable to investigate whether the advantages of this clearcut separation 
carries on to the quantum computing scenario.

At a superficial glance, the prospects of using a quantum computer for CFD look promising.
The number of qubits, $q$, required to store $Re^3$ dynamic degrees of freedom is 
simply given by:
\begin{equation}
q = 3 log_2 Re \sim 10 log_{10} Re
\end{equation}
This shows that $q \sim 70$ qubits are in principle sufficient to quantum-simulate a car.
And even numerical weather forecast, say $Re \sim 10^{10}$ could be quantum-simulated
with about $q \sim 100$ qubits, well within the {\it nominal} 
capability of current quantum hardware~\cite{bravyi_future_2022}.
 
As mentioned above, many hurdles stand in the way of this blue-sky scenario.
Leaving aside the notorious issues of quantum noise and decoherence, in the following we
focus on two issues which are specific to classical fluids: nonlinearity and dissipation.    

Indeed, the dynamics described by Eqs.~\eqref{eq:continuity_equation},\eqref{eq:velocity_equation} is nonlinear and subject to dissipation, whereas quantum algorithms consist into the application of a sequence of unitary, hence conservative, operators. This fundamental difference represents a serious obstacle to the formulation of a 
quantum algorithm capturing the NSE dynamics. 

A possible solution is provided by Carleman Linearization (CL)~\cite{carleman_application_1932,kowalski_nonlinear_1991,itani_analysis_2022}. 
This is a general strategy to transform a non-linear equation into an infinite set of linear equations, promoting all the different monomials appearing in the nonlinear equation to independent variables. In order to numerically deal with the infinite newly-defined variables, a truncation is applied at a given problem-dependent level. 

CL removes the first obstacle to the resolution of the NSE through quantum computers, non-linearity. 
Better said, it trades nonlinearity for extra-dimensions and nonlocality, as it will become apparent shortly.
To deal with the non-conservative part, we can use an extended circuit, 
first proposed in~\cite{mezzacapo_quantum_2015}, that 
makes use of an ancilla qubit to mimic the dissipative dynamics. 

Various Carleman-based LBM schemes have been proposed in the recent literature~\cite{liu_efficient_2021,li_potential_2023,itani_quantum_2023}, but, to the best of
our knowledge, none of them provided an explicit description of the corresponding quantum circuit.

\section{The lattice Boltzmann method}\label{sec:III}

The LBM has been proved particularly efficient to simulate the NSE, both at low and high Reynolds number~\cite{tran_lattice_2022}. 
LBM models the system in a $d$ dimensional regular lattice and defines 
$Q$ vectorial velocities $c_i$ pointing toward the neighboring sites. 
The probability distribution functions $f_i(x,t)$, with $i=0,\dots,Q-1$ represent the 
probability of a  representative fluid particle at lattice site $x$ at time $t$ of 
having velocity $c_i$. The vector notation is relaxed for simplicity.

The distribution functions are related to the macroscopic quantities fluid density 
$\rho$ and fluid velocity $u$ by the linear relations:

\begin{eqnarray}
\rho(x,t)&=& \sum_{i=0}^{Q-1}f_i(x,t),\label{eq:density_lbm}
\\
\rho(x,t)u(x,t)&=& \sum_{i=0}^{Q-1} c_if_i(x,t)\label{eq:velocity_lbm},
\end{eqnarray}

\begin{figure}
\centering
\includegraphics[scale=0.2]{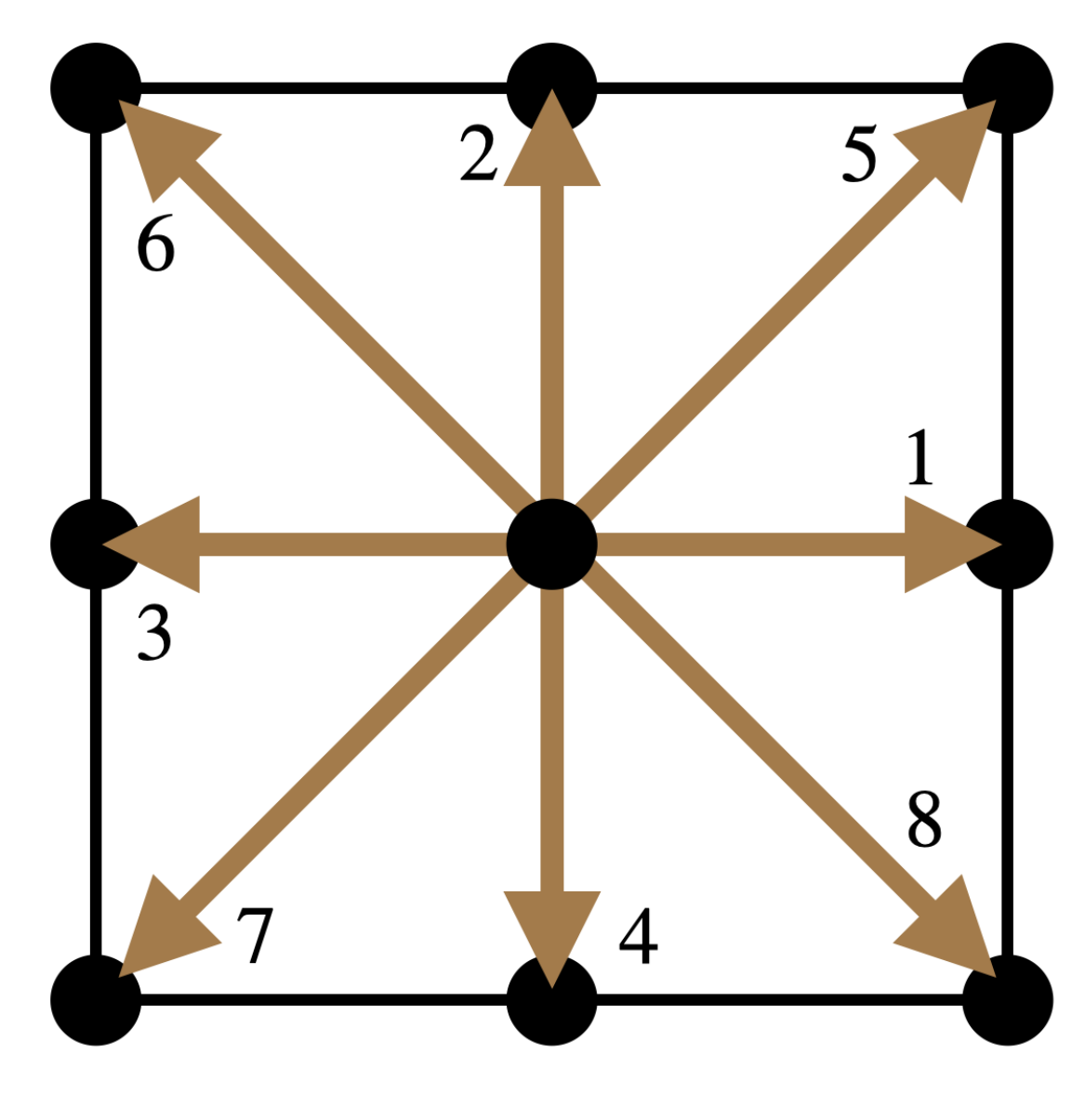}
\caption{The D2Q9 model. The model is defined on a two-dimensional square lattice, with the corresponding set 
of  nine velocity vectors labeled by index, cf. Table~\ref{tab:D2Q9}.\label{fig:D2Q9}}
\end{figure}
\begin{table}
\begin{center}
\begin{tabular}{l | c | c | c | c | c | c | c | c | c |} 
 \hline
i & $0$ & $1$ & $2$ & $3$& $4$ & $5$ & $6$ & $7$& $8$\\
 \hline 
 \hline 
 $w_i$ &  $4/9$ & $1/9$ &$1/9$ &$1/9$ &$1/9$ &$1/36$&$1/36$&$1/36$&$1/36$\\
 $c_i$ &  $(0,0)$ &$(1,0)$&$(0,1)$&$(-1,0)$&$(0,-1)$&$(1,1)$&$(-1,1)$&$(-1,-1)$&$(1,-1)$\\
\hline
\end{tabular}
\end{center}
\caption{The weights and velocity set of the D2Q9 model.\label{tab:D2Q9}}
\end{table}

\noindent where the discretized velocities $c_i$ are vectors with components either $-1,0$ or $1$, see Table~\ref{tab:D2Q9} and Fig.~\ref{fig:D2Q9} for an explicit example in $D=2$ and $Q=9$.

The lattice Boltzmann equation (LBE) reads as follows\cite{succi_lattice_2001, aidun_lattice-boltzmann_2010}:

\begin{equation}\label{eq:Boltzmann_equation}
f(x+c_i,t+1)-f_i(x,t) = -\Omega_i 
\quad \text{for }i=0,\dots, Q-1.
\end{equation}
where the left hand side is the free streaming along the $i$-th direction and the
left hand side is the discrete-velocity collision operator. The time step has made unity for simplicity.

It proves expedient to use the Bhatnagar--Gross--Krook (BGK) relaxation
expression of the collision term 

\begin{equation}\label{eq:BGK}
\Omega_i=-\frac{1}{\tau}(f_i-f_i^{\text{eq}})
\end{equation} 
where $\tau$ is the relaxation time-scale and the local equilibrium $f_i^{\text{eq}}$ is defined 
through a Taylor expansion of the Boltzmann equilibrium distribution, as follows:

\begin{equation}\label{eq:equilibrium_lattice}
f_i^{\text{eq}}(x,t) = w_i\rho(x,t)\bigg{(}1+\frac{u\cdot c_i}{c_s^2}+
\frac{(u\cdot c_i)^2}{2c_s^4}-\frac{u\cdot u}{2c_s^2}\bigg{)},
\end{equation}

\noindent where $c_s$ is the lattice speed of sound, typically $1/\sqrt{3}$ for most lattices. The weights $w_i$ can be obtained from the expansion of the equilibrium function~\eqref{eq:equilibrium_lattice} in terms of Hermite polynomials and depend on the number of dimensions $D$ and discrete velocities $Q$. 
They are reported in Table~\ref{tab:D2Q9} for the D2Q9 model.

The dynamics ruled by Eq.~\eqref{eq:Boltzmann_equation} consists of two computational steps. 
First, the collision step is given by a local and non-linear operation that transforms
the pre-collisional state $f_i(x,t)$ into the post-collisional one $f^*_i(x,t)$, as follows:

\begin{equation}\label{eq:collision_step}
f^*_i(x,t)=(1-\omega)f_i(x,t)+\omega f_i^{\text{eq}},
\end{equation}

\noindent where we have used the BGK form of the collision operator, and defined $\omega=\frac{\Delta t}{\tau}$, with $\Delta t$ being the time step of the evolution. From Eq.~\eqref{eq:collision_step} we see that if $\omega=1$ the 
system collapses to the local equilibrium at each time step. 
In general, LBM allows values of $\omega<2$ for matter of (linear) stability. 

\noindent Second, the streaming step shifts the density functions to the nearest-neighbor site, as

\begin{equation}\label{eq:streaming_step}
f_i(x+c_i,t+\Delta t)=f^*_i(x,t).
\end{equation}

\noindent The streaming process represents the free-motion of the of the fluid parcel 
across the lattice. Notice that the particles always land on a lattice site.

\subsection{Weakly compressible limit}

For weakly-compressible flows, the density can taken as nearly 
constant, with a value $\rho\approx 1$. 
This allows to write $\frac{1}{\rho}\sim(2-\rho)$, which permits to write  
the equilibrium function~\eqref{eq:equilibrium_lattice} as a 
cubic function of the density distributions, namely:

\begin{eqnarray}\label{eq:equilibrium_weakly}
f_i^{\text{eq}} & = & L_{ij}f_j+Q_{ijk}f_jf_k+T_{ijkl}f_jf_kf_l,
\end{eqnarray}

\noindent with the linear, quadratic and cubic operators defined as

\begin{eqnarray}
L_{ij} & = & w_i\bigg{(}1+\frac{c_i\cdot c_j}{c_s^2}\bigg{)},\\
Q_{ijk} & = & \frac{w_i}{c_s^4}(c_i\cdot c_jc_i\cdot c_k-c_s^2c_j\cdot c_k),\nonumber\\
T_{ijkl} & = & -\frac{1}{2}Q_{ijk},\quad \forall l.\nonumber
\end{eqnarray}

The collision step can then be rewritten in mode-coupling form as follows:

\begin{eqnarray}\label{eq:collision_weakly}
f_i^{*} & = & A_{ij}f_j+B_{ijk}f_{j}f_{k}+C_{ijkl}f_{j}f_{k}f_{l},
\end{eqnarray}

\noindent where the matrices $A, B$ and $C$ are obtained from the previous operators via the relations

\begin{eqnarray}\label{eq:weakly_AB}
A_{ij} & = & (1-\omega)\delta_{ij}+\omega L_{ij},\\
B_{ijk} & = & \omega Q_{ijk},\nonumber\\
C_{ijkl} & = & -\frac{\omega}{2}Q_{ijk},\quad \forall l.\nonumber
\end{eqnarray}

To be noted that the nonlinearity is formally carried out by the Mach 
number $\mathbf{Ma}=u/c_s$ which is $O(1)$, in stark contrast with the
Reynolds number, easily in the multimillions for macroscopic
objects, e.g. a standard automobile.
This is a potentially major advantage over the continuum formulation of
fluid dynamics. 

\hfill
\section{The Carleman linearization}\label{sec:IV}

\begin{figure}
\centering
\includegraphics[scale=0.4]{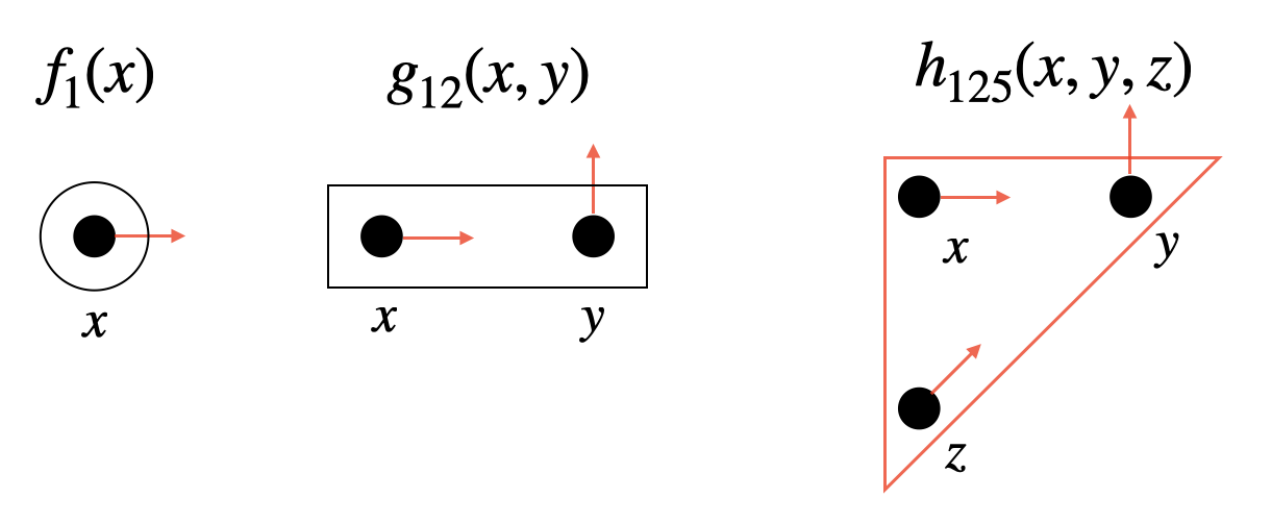}
\caption{An example of the Carleman variables of first, second and third order $f, g$ and $h$ respectively. 
In the above, $f_1(x)$ is the local variable in $x$ with velocity $c_1$, $g_{12}(x,y)=f_1(x)f_2(y)$ is 
the non-local variable of the second order with velocities $c_1$ at lattice site $x$ and $c_2$ at lattice site $y$. 
Finally, $h_{125}(x,y,z) = f_1(x)f_2(y)f_5(z)$ is the three-point non-local variable obtained by multiplication of the three functions located in $x,y$ and $z$ with velocities $c_1,c_2$ and $c_5$, respectively.
\label{fig:Carleman_variables}}
\end{figure}

As noted earlier on, the Carleman linearization~\cite{carleman_application_1932}  transforms 
a finite-dimensional non-linear problem into an infinite set of linear equations.
This technique makes the problem more suited to quantum 
computers, as the quantum mechanics of a closed system relies 
on linear algebra (leaving aside the measurement problem, of course).
It consists of assigning the status of independent dynamic
variable to any of the monomials in Eq.~\eqref{eq:collision_weakly}, thus defining the non-local variables $g_{ij}(x_1,x_2) \equiv f_i(x_1)f_j(x_2)$, $h_{ijk}(x_1,x_2,x_3) \equiv f_i(x_1)f_j(x_2)f_k(x_3)$, and so 
on for the higher degrees polynomials. 
We sketch an example of these functions in Figure~\ref{fig:Carleman_variables}.

We then define the Carleman vector $V=(f_0(x_1),\dots,g_{00}(x_1,x_1),\dots,h_{Q-1,Q-1,Q-1}(x_N,x_N,x_N),\dots)$ as the vector including all the possible products of functions $f$ localized at each of the $N$ lattice sites. 
The Carleman system is an infinite-dimensional set of linear equations, which
can symbolically be written as:

\begin{eqnarray}\label{eq:Carleman_system}
V^*=\mathcal{C}V,
\end{eqnarray}

\noindent where $V^*$ is the Carleman vector after collision, and $\mathcal{C}$ is the Carleman matrix, whose 
components can be obtained by Eq.~\eqref{eq:collision_weakly}.

\noindent In the following, we consider two different procedures for dealing with the infinite 
number of Carleman  variables, truncation and closure.  

\subsection{Carleman truncation}

We ignore the higher degree terms by applying a truncation to the Carleman system, by simply
neglecting the terms with degree above the order of the truncation. 

\subsubsection{Second order Carleman truncation}

At 2nd order, all the terms with degree $3$ or higher are set to zero, ($h=0$). 
Thus, we write: 

\begin{eqnarray}\label{eq:collision_truncation_2}
f_i^{*}(x_1) & = & A_{ij}f_j(x_1)+B_{ijk}g_{jk}(x_1,x_1),\\
g_{ij}^{*}(x_1,x_2) & = & A_{ik}A_{jl}g_{kl}(x_1,x_2).\nonumber
\end{eqnarray}

Therefore, the Carleman system can be written in terms of the vector $V_{\text{tr}}^{(2)}$, which collects all the components of $f$ and $g$ at each lattice site, and the corresponding vector after collision $V_{\text{tr}}^{(2)*}$. 
The linear relation between the two is given by the matrix $\mathcal{C}_{\text{tr}}^{(2)}$ that collects 
the elements from Eq.~\eqref{eq:collision_truncation_2}.

\subsubsection{Third order Carleman truncation}

With the truncation at 3rd order, the Carleman equations read as follows:

\begin{eqnarray}\label{eq:collision_truncation_3}
f_i^{*}(x_1) & = & A_{ij}f_j(x_1)+B_{ijk}g_{jk}(x_1,x_1)+C_{ijkl}h_{jkl}(x_1,x_1,x_1),\\
g_{ij}^{*}(x_1,x_2) & = & A_{ik}A_{jl}g_{kl}(x_1,x_2)+
A_{ik}B_{jlm}h_{klm}(x_1,x_2,x_2)+B_{ikl}A_{jm}h_{klm}(x_1,x_1,x_2),\nonumber\\
h_{ijk}^{*}(x_1,x_2,x_3) & = & A_{il}A_{jm}A_{kn}h_{lmn}(x_1,x_2,x_3).\nonumber
\end{eqnarray}

\subsection{Carleman closure}

We propose an alternative and slightly more accurate method to cut down the number of Carleman variables. 
We approximate the product of $d$ functions $f$ to the product of $d-1$ functions multiplied by a constant. 
We choose this constant to be the LBM weight of the corresponding index that has been removed, thereby
approximating the function with its steady equilibrium value.

\subsubsection{Second order Carleman closure}

We give here an explicit example of the closure procedure at second order approximating 
the following third order polynomial as

\begin{eqnarray}\label{eq:closure_2}
f_i(x_1)f_j(x_2)f_k(x_3) &=& \frac{1}{3}[f_i(x_1)g_{jk}(x_2,x_3)+f_j(x_2)g_{ik}(x_1,x_3)+f_k(x_3)g_{ij}(x_1,x_2)]\nonumber\\
&\approx& \frac{1}{3}[w_ig_{jk}(x_2,x_3)+w_jg_{ik}(x_1,x_3)+w_kg_{ij}(x_1,x_2)].
\end{eqnarray}

\noindent At this order, the closure affects directly the  definition of the 
equilibrium function, as the term $C_{ijkl}f_jf_kf_l$ changes and the closure leads to  

\begin{eqnarray}
f_i^{*} & = & A_{ij}f_j+B_{ijk}f_jf_k+C_{ijkl}f_jf_kf_l\nonumber\\
& = & A_{ij}f_j+B_{ijk}f_jf_k+\frac{1}{3}C_{ijkl}\bigg{(}w_jf_kf_l+w_kf_jf_l+w_lf_jf_k\bigg{)}\nonumber\\
& = & A_{ij}f_j+\frac{5}{6}B_{ijk}f_jf_k,
\end{eqnarray}

\noindent where the last line has been obtained via the relations

\begin{eqnarray}
C_{ijkl}w_j & = & C_{ijkl}w_k = 0, \nonumber\\
C_{ijkl}w_l & = &-\frac{B_{ijk}}{2}.
\end{eqnarray}

The Carleman form of the LBE after closure at second order is given by:

\begin{eqnarray}
f_i^*(x_1) & = & A_{ij}f_j(x_1)+\frac{5}{6}B_{ijk}g_{jk}(x_1,x_1)\\
g_{ij}^*(x_1,x_2) & = & A_{ik}A_{jl}g_{kl}(x_1,x_2)+\frac{5}{18}[w_iB_{jkl}g_{kl}(x_2,x_2)+w_jB_{ikl}g_{kl}(x_1,x_1)].
\end{eqnarray}

We see that the closure makes the collision step {\it{non-diagonal}}, as it involves $g$ functions located at different sites.

\subsubsection{Third order Carleman closure}

Closure at third or higher order does not change the equilibrium function. 
The combination of four distribution functions is approximated to

\begin{eqnarray}
f_{i}(x_1)f_j(x_2)f_k(x_3)f_l(x_4)&\approx&\frac{1}{4}[w_ih_{jkl}(x_2,x_3,x_4)+w_jh_{ikl}(x_1,x_3,x_4)+w_kh_{ijl}(x_1,x_2,x_4)\nonumber\\
&&+w_lh_{ijk}(x_1,x_2,x_3)].
\end{eqnarray}

The collision step becomes:

\begin{eqnarray}\label{eq:collision_closure_2}
f_i^{*}(x_1) & = & A_{ij}f_j(x_1)+B_{ijk}g_{jk}(x_1,x_1)+C_{ijkl}h_{jkl}(x_1,x_1,x_1),\nonumber\\
g_{ij}^{*}(x_1,x_2) & = & A_{ik}A_{jl}g_{kl}(x_1,x_2)+A_{ik}B_{jlm}\bigg{(}\frac{7}{8}h_{klm}(x_1,x_2,x_2)+\frac{1}{4}w_i\sum_nh_{lmn}(x_2,x_2,x_2)\bigg{)}\nonumber\\
&&+B_{ikl}A_{jm}\bigg{(}\frac{7}{8}h_{klm}(x_1,x_1,x_2)+\frac{1}{4}w_j\sum_nh_{lmn}(x_1,x_1,x_1)\bigg{)}\nonumber\\
h_{ijk}^{*}(x_1,x_2,x_3) & = & A_{il}A_{jm}A_{kn}h_{lmn}(x_1,x_2,x_3)+\nonumber\\
&&\frac{1}{4}[w_iA_{jl}B_{kmn}h_{lmn}(x_2,x_3,x_3)+w_{j}A_{il}B_{kmn}h_{lmn}(x_1,x_3,x_3)]\nonumber\\
&&+\frac{1}{4}[w_iB_{jlm}A_{kn}h_{lmn}(x_2,x_2,x_3)+w_{k}A_{il}B_{jmn}h_{lmn}(x_1,x_2,x_2)]\nonumber\\
&&+\frac{1}{4}[w_jB_{ilm}A_{kn}h_{lmn}(x_1,x_1,x_3)+w_{k}B_{ilm}A_{jn}h_{lmn}(x_1,x_1,x_2)]\nonumber\\
&&+\frac{1}{32}\sum_n\bigg{(}w_iw_jB_{klm}h_{lmn}(x_3,x_3,x_3)\nonumber\\
&&+w_iw_kB_{jlm}h_{lmn}(x_2,x_2,x_2)+w_jw_kB_{ilm}h_{lmn}(x_1,x_1,x_1)\bigg{)}\nonumber.
\end{eqnarray}

\section{Comparison between the exact LBM and the Carleman linearized model}\label{sec:V}

In this section, we present the simulations of a two dimensional system with constant pressure 
and no external forces. We consider a Kolmogorov-like flow on a grid of $N=N_xN_y$ points. 
The distribution functions are initialized as follows:

\begin{eqnarray}
f_i(x,y) = w_i\bigg{[}1+A_x\cos\bigg{(}\frac{2\pi}{N_y}k_x y\bigg{)} c_i\cdot c_1+A_y\cos\bigg{(}\frac{2\pi}{N_x}k_y x\bigg{)} c_i\cdot c_2\bigg{]},
\end{eqnarray}

\noindent where the wave numbers $k_{x,y}$ are integers and $A_{x,y}$ is a positive amplitude between $0$ and $1$. 

\begin{figure}[ht!]
\centering
\includegraphics[scale=0.33]{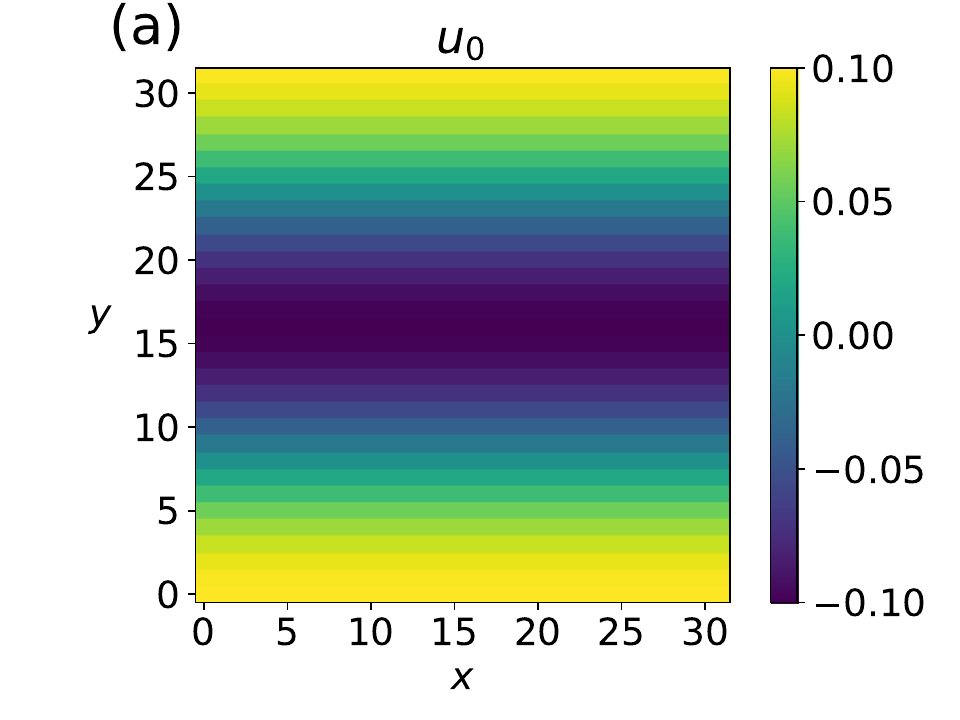}
\includegraphics[scale=0.33]{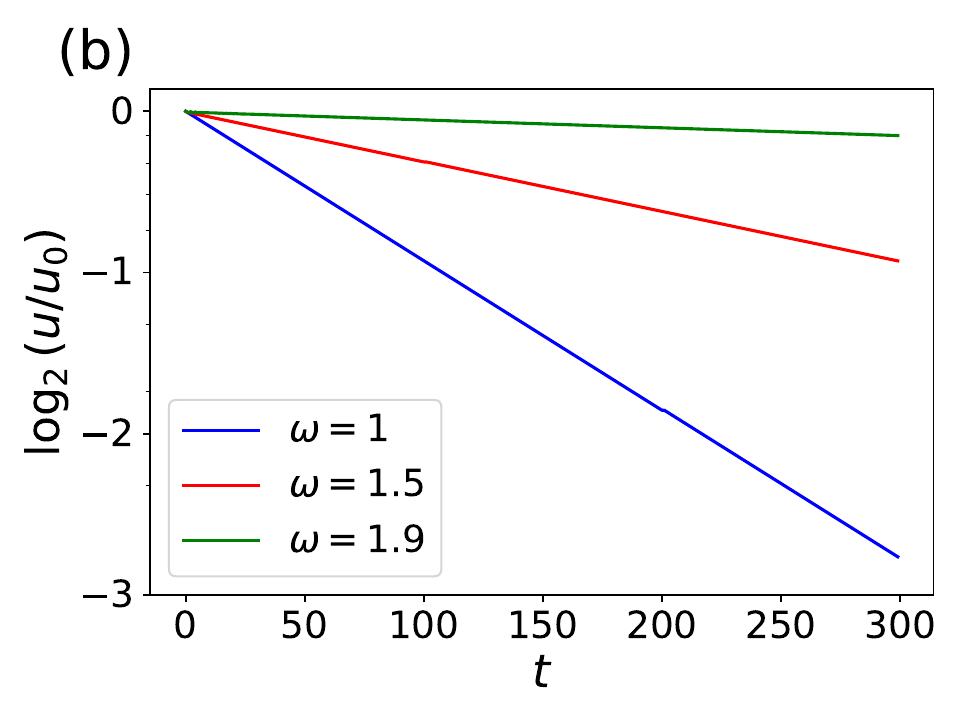}
\includegraphics[scale=0.33]{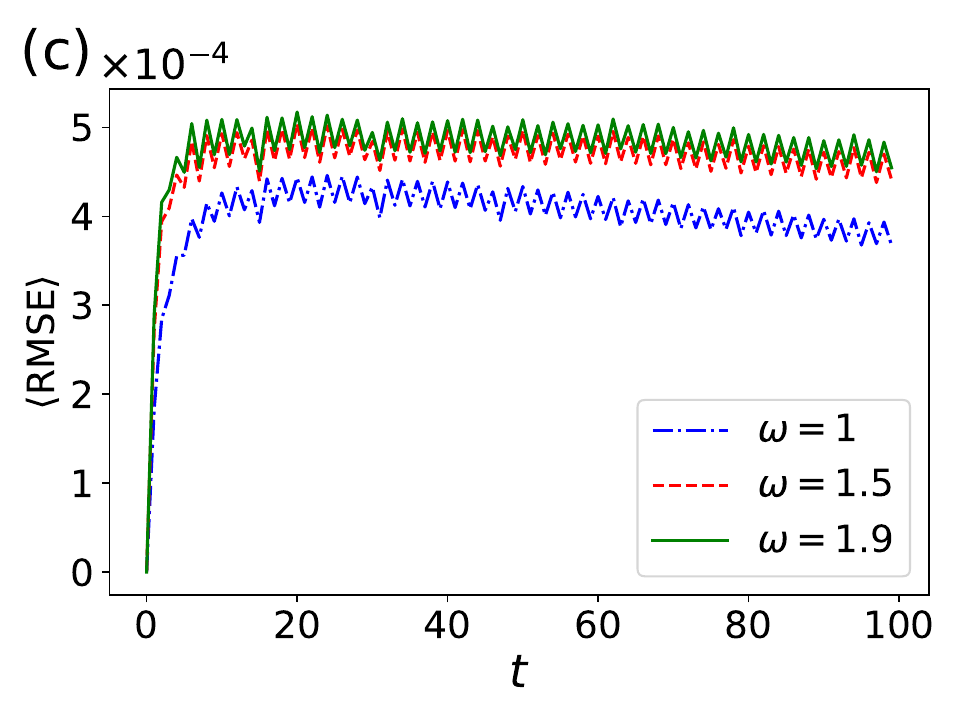}
\caption{In~(a) the initial conditions of the Kolmogorov flow on a grid with $32\times32$ grid points, with $k_x=1, A_x=0.3$. In~(b) The exponential decay of the macroscopic quantity $u_x$ for different choices of $\omega$, the exponent of the decay follows Eq.~\eqref{eq:viscosity_omega}. In~(c) the relative root mean squared error defined in Eq.~eqref{eq:RRMSE} between the results obtained by LBM and by Carleman linearization truncated at second order.  \label{fig:dissipative_kolmogorov}}
\end{figure}

By setting $A_y=0$, the velocity in the $y$ direction is null, $u_y=0$ and the dynamics is purely 
linear and dissipative, as the convective term vanishes, $u\cdot\nabla u=0$. 
The dynamics is then ruled only by the linear term $\nu\nabla^2 u$. In this regime, the velocity $u_{x}$ evolves following an exponential decaying function, with $u_x(t) = u_x(0)\exp\{-\nu k^2 t\}$, with the 
viscosity $\nu$ being a function of the LBM parameter $\omega$,

\begin{equation}\label{eq:viscosity_omega}
\nu = \frac{1}{6}\bigg{(}\frac{2}{\omega}-1\bigg{)},
\end{equation}

\noindent in lattice units with $\Delta t = \Delta x =1$.
For this linear regime, we show in Fig.~\ref{fig:dissipative_kolmogorov}(a) the velocity at $t=0$ for a grid with $N=32\times 32$, choosing $A_k=0.3$ ($u_x=0.1$) and $k_x=1$. Simulations using the ``exact" LBM 
recover the exponential decay, as shown in Fig.~\ref{fig:dissipative_kolmogorov}(b) for different 
values of $\omega$.

In order to compare the LBM with the CL approximation, we define the 
Root Mean Squared Error (RMSE) $\epsilon(a,b)$ between two distributions $a$ and $b$ as

\begin{equation}\label{eq:epsilon_RRMSE}
\epsilon(a,b) = \sqrt{\sum_{i=1}^{N}\frac{1}{N}\bigg{(}\frac{a_i-b_i}{a_i}\bigg{)}^2}.
\end{equation}

In the above, $N$ is the number of grid points. 
Accordingly, we end up with one RMSE 
for each of the $Q$ velocity directions of the LBM. 
In our analysis, we consider the mean value of the RMSE among the $Q$ distributions, namely:

\begin{equation}\label{eq:RRMSE}
\langle{\text{RMSE}}\rangle = \sum_{q=0}^{Q-1}\frac{1}{Q}\epsilon(f_q^{\text{LBM}},f_q^{\text{CL}}),
\end{equation}

\noindent where $f_q^{\text{LBM}},f_q^{\text{CL}}$ are the distribution functions calculated with the LBM and the CL respectively. The quantity $\langle{\text{RMSE}}\rangle$ accounts for the mean deviation of the CL with respect to the results obtained by the LBM. From Fig.~\ref{fig:dissipative_kolmogorov}(c), we see that this deviation is of the order of $10^{-4}$.
This is fully consistent with the weak compressibility error of the standard LBM. 

\begin{figure}[ht!]
\centering
\includegraphics[scale=0.33]{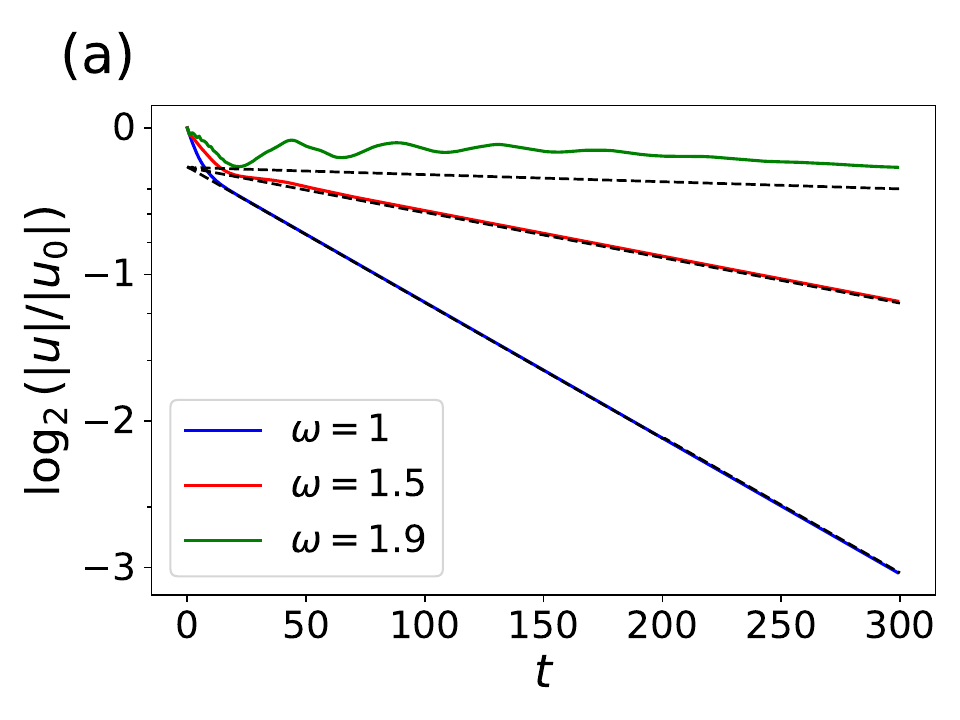}
\includegraphics[scale=0.33]{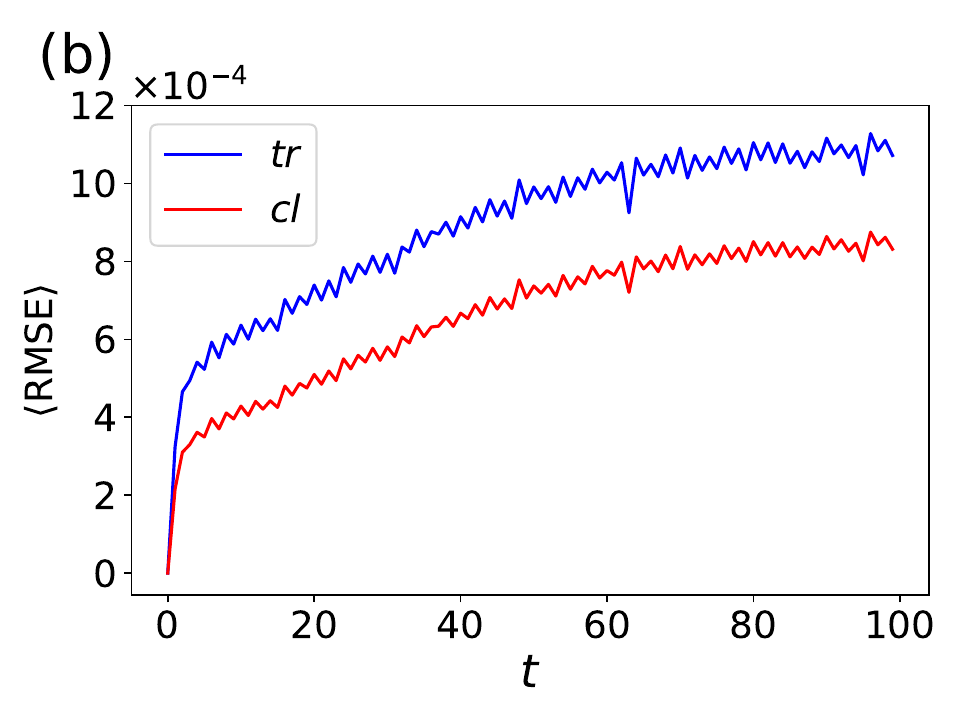}
\includegraphics[scale=0.33]{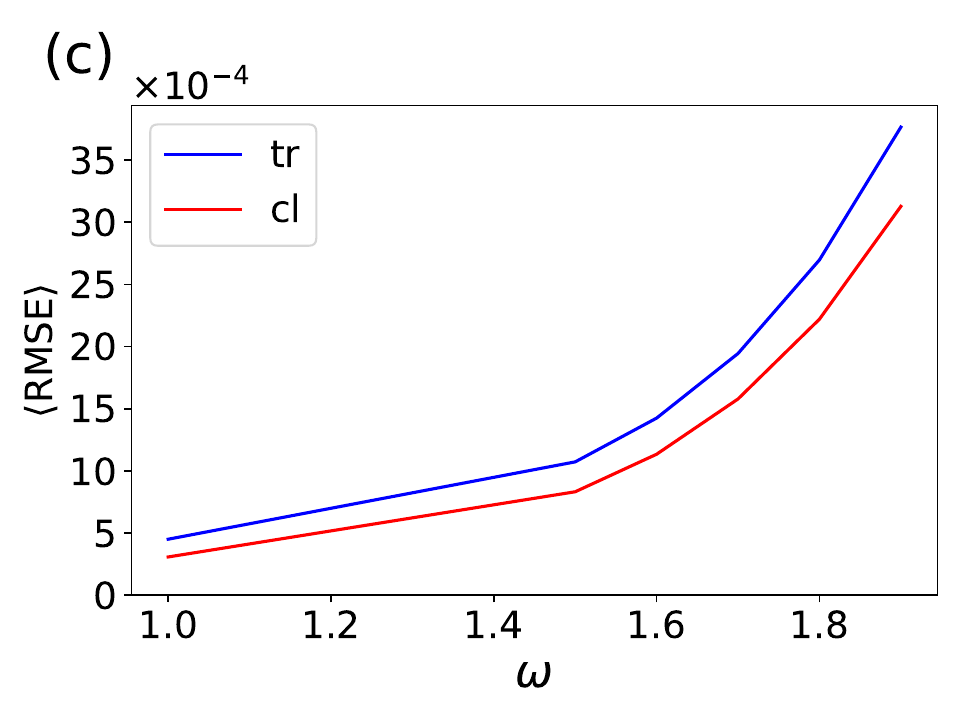}
\caption{(a) The temporal evolution of $u_x$ for different values of $\omega$ with initial conditions $A_x=0.3$,$A_y=0.2$, $k_x=1$, and $k_y=4$, (b) the comparison between the $\langle{\text{RMSE}}\rangle$ of the truncation method (with label $\text{'tr'}$) and the $\langle{\text{RMSE}}\rangle$ of the closure method (with label $\text{'cl'}$) for the Kolmogorov flow, varying $\omega$, and (c) the comparison between the $\langle{\text{RMSE}}\rangle$ of the truncation and the closure methods at $t=100$ as a function of $\omega$ . \label{fig:kolmogorov_error_omega}}
\end{figure}

Upon increasing the value of $A_y$, we start observing the effects of the non-linearity, as the convective terms raises in magnitude and the Reynolds number becomes larger. 
Thus, the temporal evolution of $u_x$ and $u_y$ deviates from the exponential decay, as it is shown in Figure~\ref{fig:kolmogorov_error_omega}(a), with parameters $A_x=0.3$, $k_x=1$, $A_y=0.2$, $k_y=4$. 
For low $\omega$ (high dissipation, low Reynolds) 
the curves rapidly go back to the original decaying dynamics, whereas for $\omega=1.9$, the deviation is evident even after longer time periods. 
The oscillations of the curve are due to the presence of small vortices caused by the mildly turbulent dynamics.

In Fig.~\ref{fig:kolmogorov_error_omega}(b) we compare the $\langle\text{RMSE}\rangle$ between the truncation and the closure at second order. The Figure refers to $\omega=1.5$, but similar deviations are obtained for different $\omega$, as it is evident from Fig.~\ref{fig:kolmogorov_error_omega}(c).
We see that the approximation brought by the closure leads to a lower $\langle{\text{RMSE}}\rangle$, thus mitigating the error. However, we notice that the order of magnitude between the two cut-off methods remains the same. 
The $\langle{\text{RMSE}}\rangle$ depends on the value of $\omega$, as shown in Fig.~\ref{fig:kolmogorov_error_omega}(c), and therefore on the corresponding value of the Reynolds number $\text{Re}$. 
At low Reynolds the $\langle{\text{RMSE}}\rangle$ remains below $10^{-3}$, which is a 
noteworthy result given the small number of grid-points employed in these simulations.

We conclude that the Carleman truncation(closure) at second order does not introduce any error
beyond the one inherent to the lattice Boltzmann procedure, at least up to $Re \sim O(100)$.
While still far from turbulence, this is nonetheless encouraging, also in view of the fact that there
are interesting and demanding problems in the physics of low-Reynolds fluids, particularly in the
context of biology and soft matter~\cite{dunweg_lattice_2009,succi_lattice_2022}, and in quark-gluon plasma hydrodynamics~\cite{mendoza_fast_2010}.

\section{Quantum circuit}\label{sec:VI}

In the previous section we have shown the performance of the CL as applied to the LBM. 
We have seen that, even truncating the number of Carleman variables 
at just second order, the error is about $10^{-3}$. 

However, the number of Carleman variables increases as $\mathcal{O}(N^kQ^k)$,  $k$ being the 
truncation order, since we need to multiply together all possible combinations of distributions functions 
with different velocity at different spatial lattice locations. 

To convey a concrete idea of the numbers in play, we observe that a simulation of the D2Q9 model 
on a $32\times 32$ square lattice implies a number of natural variables $n_v = N\times Q=9216$, a number of Carleman variables at second order $n_{\text{CL}}^{(2)}\approx9\times10^7$ and at third order $n_{\text{CL}}^{(3)}\approx8\times10^{11}$. As a touchstone, the current near-exascale supercomputers  can handle
of the order of trillion ($10^{12}$) grid points. 
Clearly, CL is not a viable option on classical computers: the dimensionality and nonlocality 
price is much higher than the linearity gain. 

One may argue that for each time step, the only non-local terms that affect the dynamics 
are the ones involving neighbor sites, which would substantially reduce the number of Carleman variables. 

\begin{figure}
\centering
\includegraphics[scale=0.2]{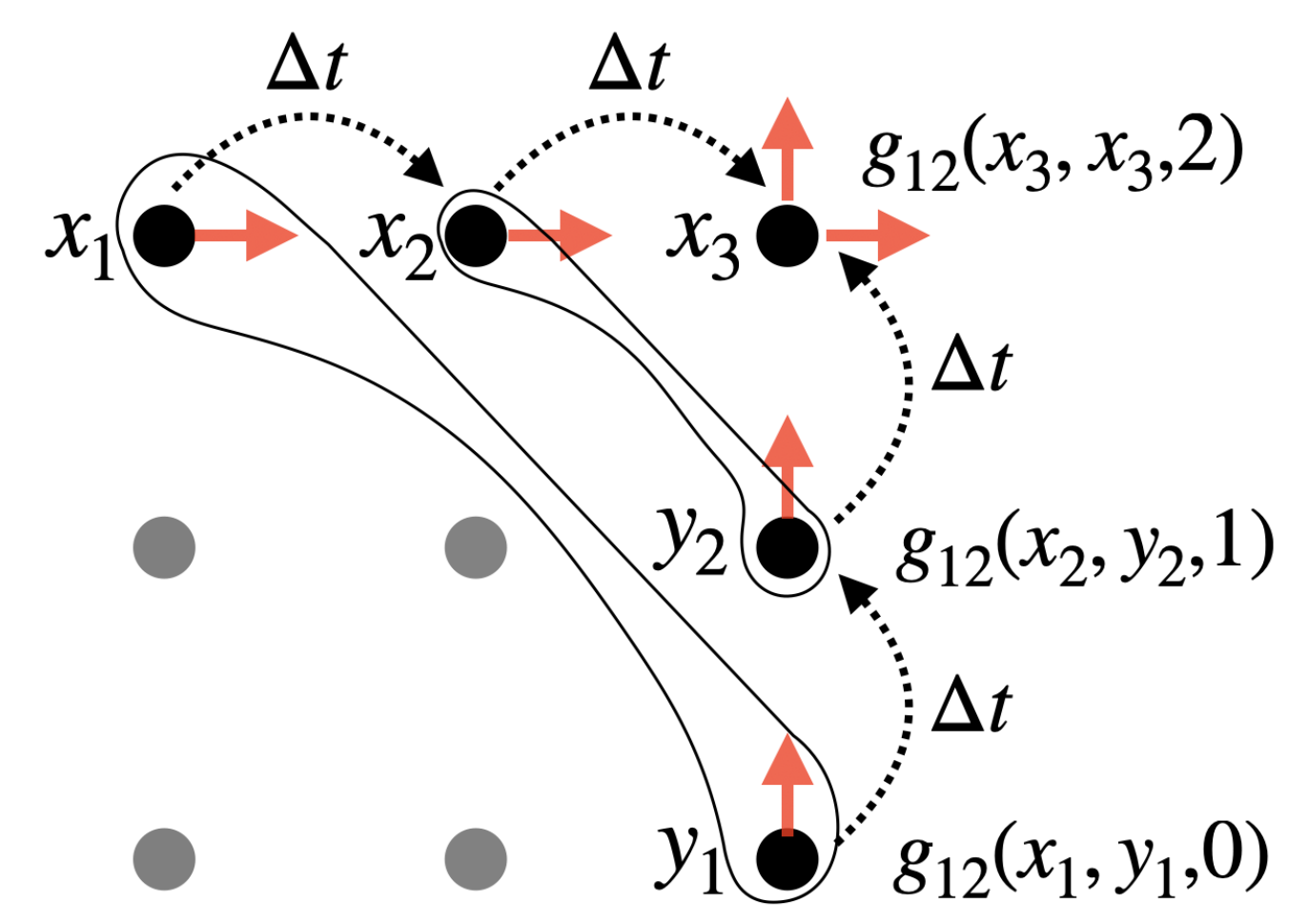}
\caption{The two-point non-local function $g(x_1,y_1,0)$ at $t=0$ streams to neighboring sites after one time step $\Delta t$, thus defining the non-local functions $g(x_2,y_2,1)$. After one more time step, the streaming lands into the local 
pair function $g(x_3,x_3,2)$ at $t=2$. \label{fig:nonlocalg}}
\end{figure}

However, non-local correlations, involving neighbors up to order $s$ 
are required,  if the simulation is run as a single update from time
$t$ to time $t+s$. 
\noindent For instance, when a non-local variable $g$ is calculated through the collision process, it streams into the neighboring sites serving as initial data for the subsequent collision process. Figure~\ref{fig:nonlocalg} illustrates the flow of the two points non-local function $g(x_1,y_1)$ until it converges at the point $x_3$, thereby defining the local function $g(x_3,x_3)$.

The numbers provided above correspond to the global case $s\geq \frac{L+1}{2}$, where 
$L$ is the linear size of the spatial domain. 
The situation changes if one aims to compute a number of time steps $s<\frac{L+1}{2}$. The number of variables reduces to $N(Q+Q^2(1+2(s-1))^2)$. For a single time step this simplifies to $n_{\text{CL}}=N(Q+Q^2)$, a significant reduction of variables.

Nevertheless, the purpose of CL is to explore whether a quantum computer could do away with the above problems, and for the application on quantum computers, it may be more convenient to deal with a high number of variables than to reset the calculation and reinitialize the quantum state~\cite{zoufal_quantum_2019}. At the same time, dealing with a single time step leads to significant simplifications,  to be detailed shortly.

Future quantum computers might be able to handle this explosive increase of variables in case of 
large time-steps simulation, as  they would need a number of qubits $q = log_2n_{\text{CL}} $. 
This translates into only about $q = 27$ for the truncation at second order, and $q = 40$ for 
the truncation at third order, both numbers being well within the 
{\it nominal} reach of current quantum computers.

\subsection{The quantum embedding} 
We can embed the whole Carleman vector with amplitude encoding in a quantum state in the following way. 
To embed the linear components $f$ we can use two quantum registers, such that 
\begin{equation}\label{eq:linear_embedding}
|f\rangle = \sum_{i=0}^{Q-1}\sum_{x=0}^{N-1}f_{i}(x)|i\rangle_{v_1}|x\rangle_{p_1},
\end{equation} 

\noindent where the subscripts ${v_1},p_1$ stand for velocity and position registers respectively. The register $v$ is composed by $\lceil \log_2Q\rceil$ qubits, while the register $p$ by $\lceil \log_2N\rceil$ qubits.
To embed the quadratic components $g$ we use four quantum registers, such that

\begin{equation}\label{eq:quadratic_embedding}
|g\rangle = \sum_{i,j=0}^{Q-1}\sum_{x,y=0}^{N-1}g_{ij}(x,y)|i\rangle_{v_1}|j\rangle_{v_2}|x\rangle_p|y\rangle_p,
\end{equation} 

and the same embedding strategy can be applied for functions of higher degree. 
In order to assemble the Carleman vector collecting together $f$ and $g$, we use an extra 
quantum register that contains the information about the truncation order $\tau$. 
This is made by $\lceil\log_2\tau\rceil$ qubits (just 1 qubit is necessary to embed 
the truncation at second order). Thus, the two vectors can be merged together
\begin{eqnarray}\label{eq:quantum_embedding}
f&\to& |f\rangle|0\rangle_{v_2}|0\rangle_{p_2}|0\rangle_\tau,\\
g&\to& |g\rangle|1\rangle_\tau.
\end{eqnarray}

Although this embedding doesn't fill all the components of the quantum state, it provides 
a helpful way to define the streaming and collision operators.  
In the remaining of this section we propose a concrete way to implement the streaming and 
collision steps of LBM with CL in terms of quantum operators.

\subsection{The Multi-streaming operator}

In this section we analyze the streaming step of the LBM and its effect on the Carleman variables. 

The linear components of the Carleman vector $V,V^*$ of equation~\eqref{eq:Carleman_system},before 
and after collision respectively, $f,f^{*}$, are $N\times Q$, where $N=N_xN_y$ is the number of lattice sites.
For these variables, the streaming is simply given by Eq.\eqref{eq:streaming_step}. 
We can define $S_i$ as the linear operator embedding the transformation to be applied 
to the linear components $f^*_i$ with velocity $c_i$. 
The streaming operation can thus be written as

\begin{equation}
f(t+\Delta t) = \bigoplus_{i=0}^{Q-1}S_if_i^{*}(t).
\end{equation}

where $\bigoplus$ is the symbol of the direct sum, i.e. each streaming operator $S_i$ acts only 
on the subspace of the functions $f_i$ and performs the shift $f_i(x,t+\Delta t) = f_i^{*}(x-c_i,t)$. 

\noindent The streaming operation $S_i$ is therefore a controlled operation, conditioned by the value of the velocity register $v_1$. The explicit form of the streaming operators $S_i$ can be obtained from an adaptation of the circuit proposed in~\cite{todorova_quantum_2020}, that uses just a polynomial number of two-qubit gates per streaming operator.

We can consistently define the vector of the second order Carleman variables before and after collision $g,g^*$, where the quadratic components are given by all the possible pairs of $Q$ and $N$, as stated before.
The streaming on the quadratic component $g_{ij}$ at position $(x_1,x_2)$ applies the transformation

\begin{equation}
g_{ij}(x_1,x_2,t+\Delta t) = g^*_{ij}(x_1-c_i,x_2-c_j,t),
\end{equation} 
and the corresponding operator is given by the tensor product of the linear streaming operators as

\begin{equation}\label{eq:bistreaming}
g(t+\Delta t) = \bigoplus_{i,j=0}^{Q-1}S_i\otimes S_j g^{*}_{ij}(t).
\end{equation}

\begin{figure}
\centering
\includegraphics[scale=0.4]{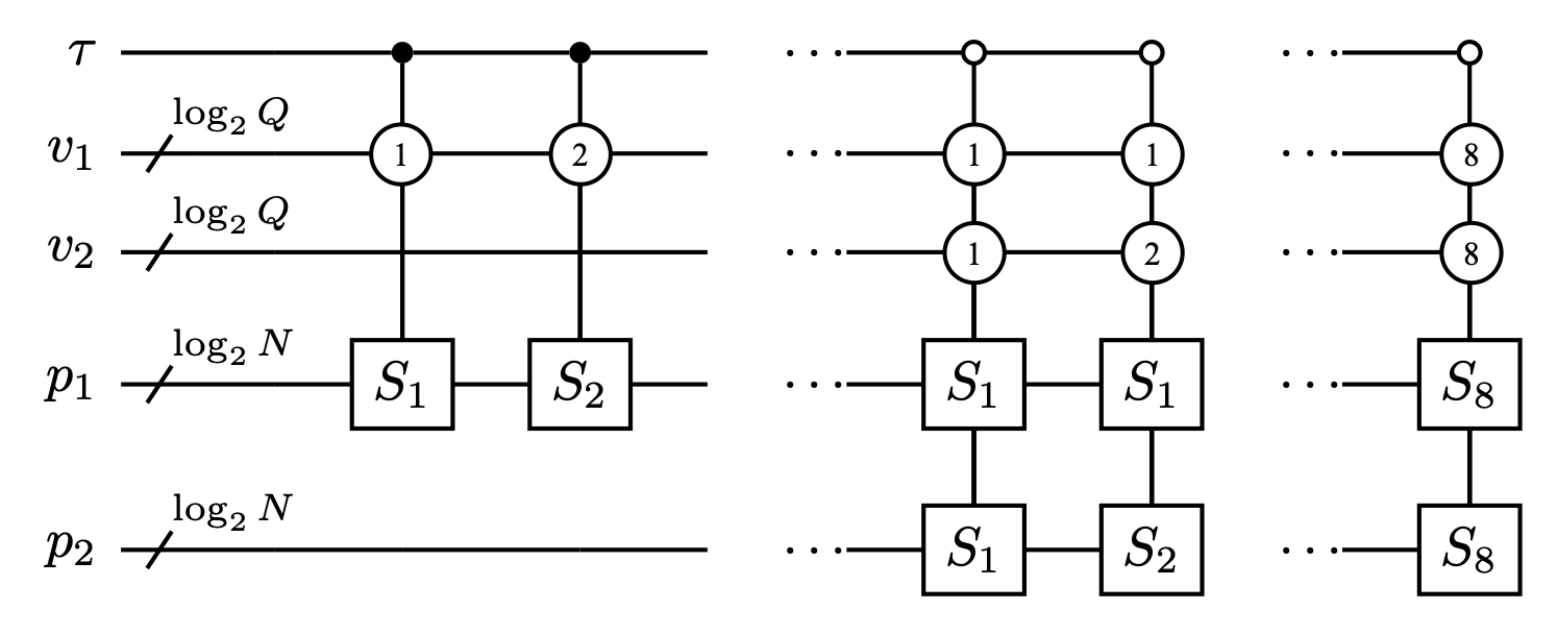}
\caption{The circuital representation of the streaming process for a Carleman truncation at second order. In this case the quantum register $\tau$ consists of just one qubit. When the value of the $\tau$ qubit is 0, the streaming operator is applied on the first position register controlled by a value of the first velocity register. When the value of the $\tau$ qubit is 1, two streaming operators are applied controlled by the values of two velocity registers. \label{fig:multistreaming}}
\end{figure}

\noindent From the circuital point of view, the streaming operator $S_i\otimes S_j$ is a double-controlled operation conditioned by the velocity registers $v_1$ and $v_2$. In fact, the tensor product in Eq.~\eqref{eq:bistreaming} means that the we can just apply linear streaming operators on the different position registers, as depicted in Fig.~\ref{fig:multistreaming}.

\begin{figure}
\centering
\includegraphics[scale=0.3]{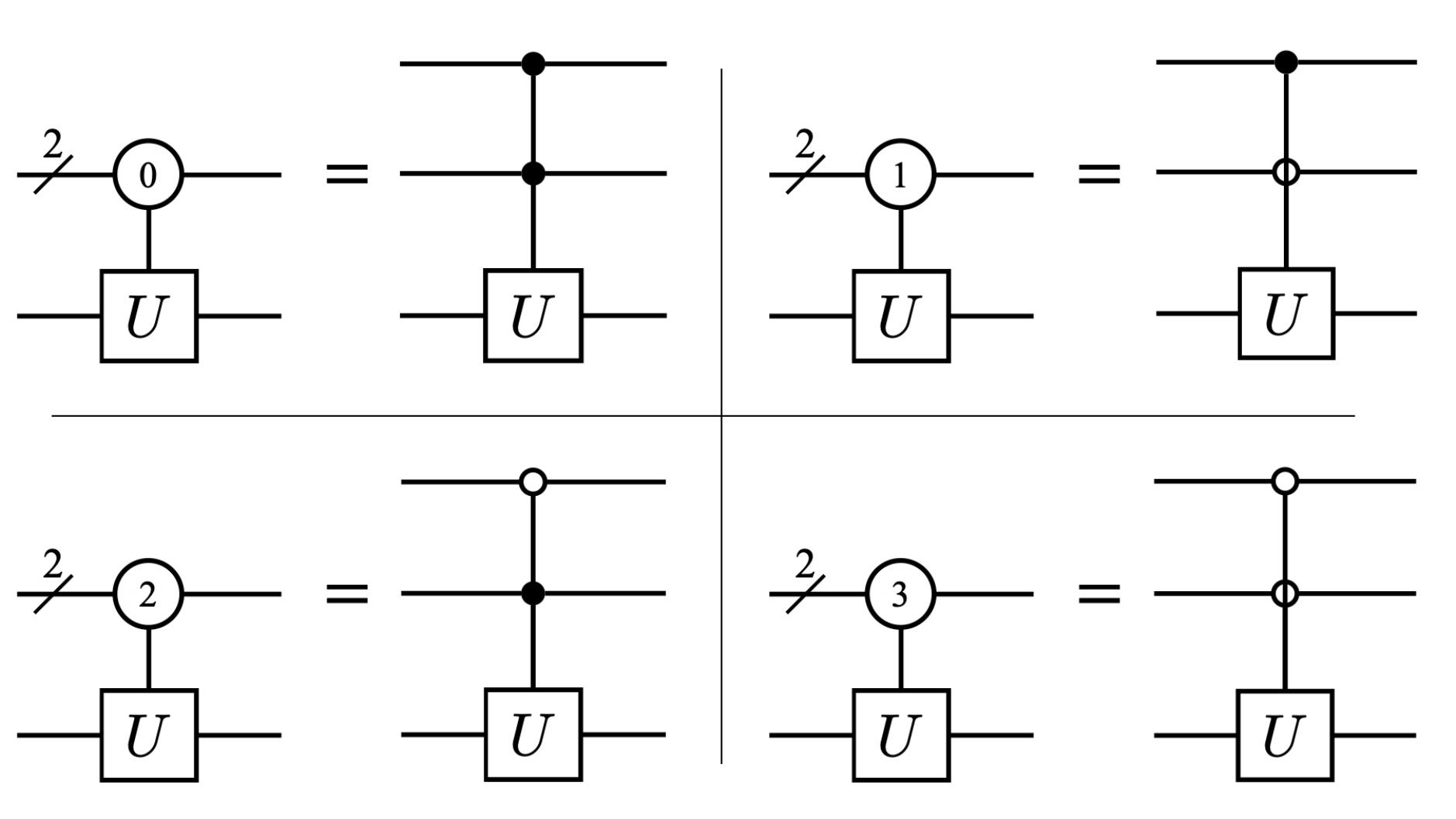}
\caption{The symbol for the multi-qubit controlled operation conditioned by the numerical value of the bit string for a simple scenario with two qubits as control.\label{fig:multiCU}}
\end{figure}

As the two velocity registers run over $\log_2Q$ qubits, we have introduced the symbol for a multi-qubit controlled operation, conditioned by the numerical value of the $Q-bit$ string. In Fig.~\ref{fig:multiCU} we explicitly define the symbol for a simple case of two qubits acting as control.


Furthermore, we see that the streaming in the diagonal direction, i.e. the one carried by the velocities $c_5,c_6,c_7$ and $c_8$, cf. Table~\ref{tab:D2Q9}, can be written as the composition 
of two streaming operators in horizontal and vertical direction, so that 

\begin{equation}
S_5 = S_1S_2,\quad S_6 = S_2S_3,\quad S_7 = S_3S_4,\quad S_8 = S_4S_1.
\end{equation}

The explicit form of the bi-streaming operator allows us to extend its representation to higher Carleman orders. 
For instance, the streaming of the cubic components of the Carleman vector is given by the 
tensor product of three streaming operators, as

\begin{equation}
h(t+\Delta t) = \bigoplus_{i,j,k=0}^{Q-1}S_i\otimes S_j\otimes S_k h_{ijk}^{*}(t),
\end{equation}

\noindent and the extension applies naturally to Carleman variables of higher order.

We notice that when performing the collision step, we need to calculate also the product of functions at 
different lattice sites, as it is explicitly apparent from Eq.~\eqref{eq:bistreaming}. 
The application of the bi-streaming operator requires a non-local collision operator, i.e. we need to 
calculate also the combinations $g^*(x_1,x_2)$ for each pair of grid points. 
This is of course a major downside of CL as a quantum computing method, one
 that complicates the circuital expression of the collision operator, as we are going to show next. 

\subsection{The collision operator}

The collision step of the LBM is a non-linear operation that implements the relaxation of the 
probability distributions towards the equilibrium distributions~\eqref{eq:equilibrium_weakly}. 
By definition, the dissipative process induced by the relaxation cannot be represented by a 
circuit performing unitary operations. 
However, we can circumvent this issue by using an ancilla qubit and making 
use of open quantum system theory. 
We can therefore implement a circuit that allows only unitary operations, but where  
only a subset of the available qubits is accounted for (the system).
By tracing out the ancilla qubits (environment), we obtain a non-unitary operation on the system qubits. 

The circuit that implements the collision operator is detailed in Ref.~\cite{mezzacapo_quantum_2015}. 
This circuit is capable of performing non-unitary operations on the qubits by means of an ancilla qubit. 
It achieves this by decomposing a non-unitary, positive-definite matrix $C$ into a linear combination 
of two unitaries $U_a$ and $U_b$, such that
 
\begin{equation}\label{eq:weighted_sum}
C=U_a+\gamma U_b.
\end{equation} 

\noindent The coefficient $\gamma$ is chosen such that the maximum eigenvalue of $C$, $c_M$, fulfills the relation $c_M\leq 1+\gamma$. Moreover, a requirement needed to apply this 
decomposition is that $c_M-c_m<2$, where $c_m$ is the minimum eigenvalue of $C$. 
Whenever this is not the case, we can define a renormalized matrix $\tilde{C}=C/c_M$ and apply the decomposition to the new matrix. The coefficient has to be set to $\gamma = 1-c_m/c_M$~\cite{mezzacapo_quantum_2015}.

\begin{figure}[t]
  \centering
\begin{quantikz}
\lstick{$|0\rangle$} & \gate{R_y(\Gamma)} & \octrl{1}  & \ctrl{1} & \gate{R_y(-\Gamma)}& \meter{}  \vcw{1}\\
\lstick{$|\psi\rangle$} & \qw & \gate{U_b} &	\gate{U_a} & \qw & \gate{S} & \qw
\end{quantikz}  
\caption{The circuit for the implementation of the collision step. The initial state $|\psi\rangle$ contains the collection of the Carleman variables. Depending on the result of the measurement on the ancilla qubit, the algorithm is successful and the streaming operator $S$ is applied.}
  \label{fig:Childs_circuit}
\end{figure}
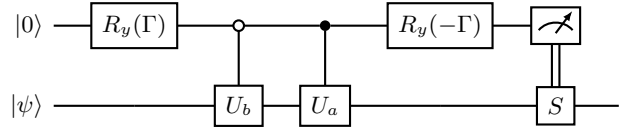

The circuit is represented in Fig.~\ref{fig:Childs_circuit}. The Carleman vector is embedded in the state $|\psi\rangle$, and an ancilla qubit is initialized in $|0\rangle_a$. 
We perform an $R_y$ rotation of angle 
$\Gamma=\arccos\bigg{(}\sqrt{\frac{\gamma}{\gamma+1}}\bigg{)}$, on the ancilla qubit, yielding the state $|\psi\rangle(\cos\Gamma|0\rangle_a + \sin\Gamma|1\rangle_a)$. The action of the anti-controlled and controlled unitaries leads to the state $U_b|\psi\rangle\cos\Gamma|0\rangle_a + U_a|\psi\rangle \sin\Gamma|1\rangle_a$ and the inverse rotation $R_y^{\dagger}(\Gamma)$ on the ancilla qubit results into


%
%

\begin{eqnarray}
\frac{1}{\gamma+1}|0\rangle(U_a+\gamma U_b)|\psi\rangle+\frac{\sqrt{\gamma}}{\gamma+1}|1\rangle(U_a-U_b)|\psi\rangle.
\end{eqnarray}

\noindent We finally measure the ancilla qubit. 
If the outcome is $|0\rangle$, the state collapses onto $(U_a+\gamma U_b)|\psi\rangle$, which is exactly the application of $C$ on the state $\psi$. In this case, the algorithm succeeds and we can proceed with the streaming step. 
On the other hand, if the outcome of the measurement is $1$, the update fails and the circuit needs to be repeated. 
The probability of success is constrained by $p_0\leq\frac{4\gamma}{(\gamma+1)^2}$.

Inspection of the collision matrix shows that it can be explicitly written as follows: 

\begin{eqnarray}
\mathcal{C} &=& |0\rangle\langle0|_{\tau}\otimes\mathds{1}_{p_1}\otimes\sum_{ij} A_{ij}|i\rangle\langle j|_{v_1}\otimes|0\rangle\langle0|_{p_2}\otimes|0\rangle\langle0|_{v_2}+\nonumber\\
& &|0\rangle\langle1|_{\tau}\otimes\mathds{1}_{p_1}\otimes\sum_{ijk} B_{ijk}|i\rangle\langle j|_{v_1}\otimes\sum_x|0\rangle\langle x|_{p_2}\otimes|0\rangle\langle k|_{v_2}+\nonumber\\
&&|1\rangle\langle1|_{\tau}\otimes\mathds{1}_{p_1}\otimes\sum_{ik} A_{ik}|i\rangle\langle k|_{v_1}\otimes\mathds{1}_{p_2}\otimes\sum_{jl} A_{jl}|j\rangle\langle l|_{v_2},
\end{eqnarray}

\noindent or, in more compact form:
\begin{equation}\label{eq:matrixC}
\mathcal{C} = \begin{pmatrix}
\mathcal{A}\otimes |0\rangle\langle0|_{p_2}\otimes |0\rangle\langle0|_{v_2} & \mathcal{B}\otimes \sum_x|0\rangle\langle x|_{p_2}\\
0 & \mathcal{A}\otimes \mathcal{A}\otimes\mathds{1}_{p_2}
\end{pmatrix}\otimes\mathds{1}_{p_1},
\end{equation}

\noindent where $\mathcal{A}$ and $\mathcal{B}$ represent the $A$ and $B$ matrices of Eq.~\eqref{eq:weakly_AB} embedded in the space of qubits. The different quadrants of the matrix~\eqref{eq:matrixC} refer to the values of the $\tau$ register. We note that all the components depend in a non-trivial way on the $p_2$ register. 
This does not permit to define the collision operator as a local process to be applied only on the velocity 
registers $v_1$ and $v_2$,  a feature which is rooted in the inherent 
non-locality of the  Carleman linearization. 
Because of this, the matrix cannot be written in sparse, block-diagonal 
form, a well known requirement for polynomial approximations in
the number of quantum gates~\cite{berry_efficient_2007,jordan_efficient_2009}. 
Thus, the two controlled operations of the unitaries $U_a,U_b$ require
a number of two-qubit gates of the order of $4^n$, according to the theoretical lower bound~\cite{barenco_elementary_1995}. 
Since the number of qubits in our system is $n=\lceil{\log_2(NQ+N^2Q^2)}\rceil+1$, for a 
Carleman system truncated at second order, the corresponding
number of two-qubit gates scales as $\mathcal{O}(N^4Q^4)$. 

For relevant cases, this number exceeds by several orders of magnitude the 
current capacity of any quantum computer~\cite{tannu_not_2019,bravyi_future_2022}. 
Just a simple $32 \times 32$ grid with $9$ discrete populations, features
$(NQ)^4 \sim 9000^4 \sim 10^{16}$ two-qubit gates.

We numerically tested this result for the circuit of 
the collision operator with the IBM \text{Qiskit} package. 
To streamline the numerical analysis, we define an hermitian augmented Carleman 
matrix $\mathcal{C}^\text{H}$ as;

\begin{equation}
\mathcal{C}^\text{H} = \begin{pmatrix}
0 & \mathcal{C}\\
\mathcal{C}^{T} & 0
\end{pmatrix}.
\end{equation}
where superscript $T$ stands for ``transpose".

\noindent 
This passage simplifies the numerical representation of $\mathcal{C}^\text{H}$ 
as the weighted sum of unitary matrices (Eq.~\eqref{eq:weighted_sum}) with the minimal 
addition of just one qubit. 
With this matrix at hand, we can translate the circuit into a sequence of two-qubit gates 
using \text{Qiskit}'s decomposition tool. 
This straightforward step gives a number of two-qubit gates close to the theoretical lower-bound 
mentioned earlier, indicating that simplifying the circuit is by no means a trivial task. 
We stress here that this issue is common to any algorithm which aims at implementing a non-sparse matrix.

\subsubsection{Single time-step collision operator}

A possible getaway from this issue is to introduce a \textit{single-step} collision operator. 
As highlighted in section~\ref{sec:VI}, the non-local Carleman variables are needed only 
if one aims to apply the dynamics over multiple time steps. 
However, if only one single time step is implemented, the number of Carleman variables 
reduces to $N(Q+Q^2)$. Since all the Carleman variables are now local, the quantum 
register $p_2$ can be dropped, and the matrix~\eqref{eq:matrixC} takes the simpler form:

\begin{equation}\label{eq:matrixC_1}
\mathcal{C}_{s=1} = \begin{pmatrix}
\mathcal{A}\otimes |0\rangle\langle0|_{v_2} & \mathcal{B}\\
0 & \mathcal{A}\otimes \mathcal{A}
\end{pmatrix}\otimes\mathds{1}_{p_1},
\end{equation}

\noindent which is also local in qubit's space. 
This means that we can apply the collision operator only on the registers 
embedding the information about truncation order $\tau$ and the velocities $v_1,v_2$.

In this framework, we can exploit the symmetry of the second order functions, 
such that $g_{ij}=g_{ji}$, to reduce even further the dimension of the number 
of Carleman variables to $\frac{3}{2}NQ+\frac{NQ^2}{2}$. Consequently, for the D2Q9 model, the matrix $\mathcal{C}_{s=1}$ of Eq.~\eqref{eq:matrixC_1} can be written with $54$ variables, and be embedded in the space of $q=6$ qubits, {\it regardless of the number of lattice sites}. 

\begin{figure}[ht!]
\centering
\includegraphics[scale=0.33]{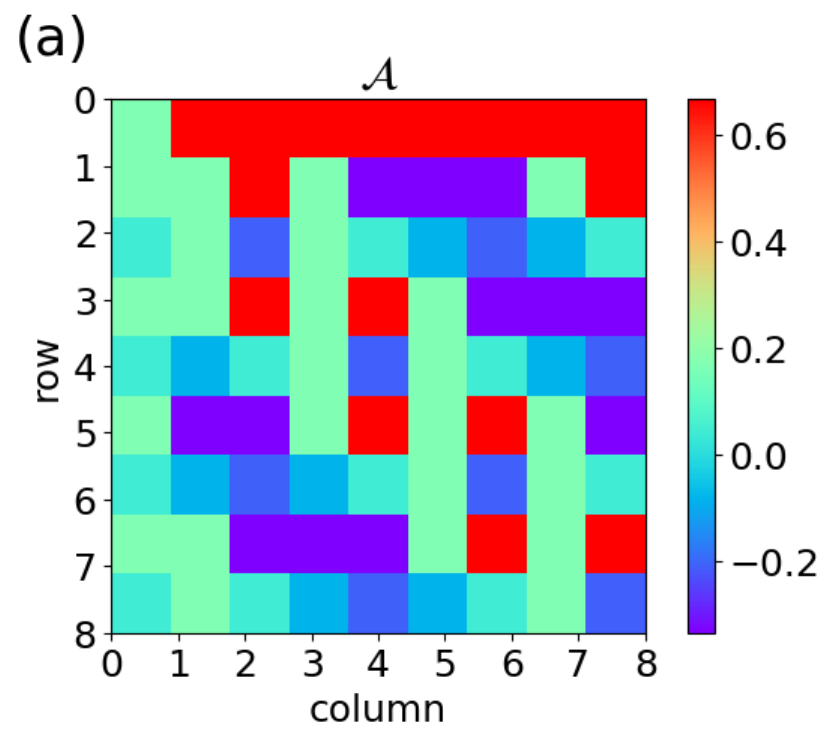}
\includegraphics[scale=0.33]{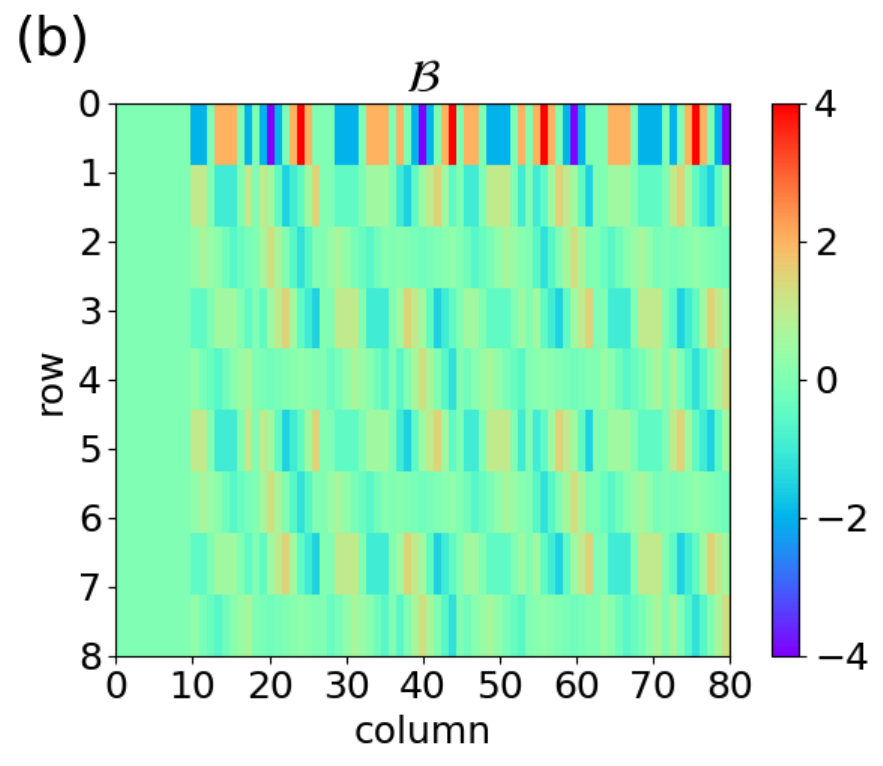}
\includegraphics[scale=0.33]{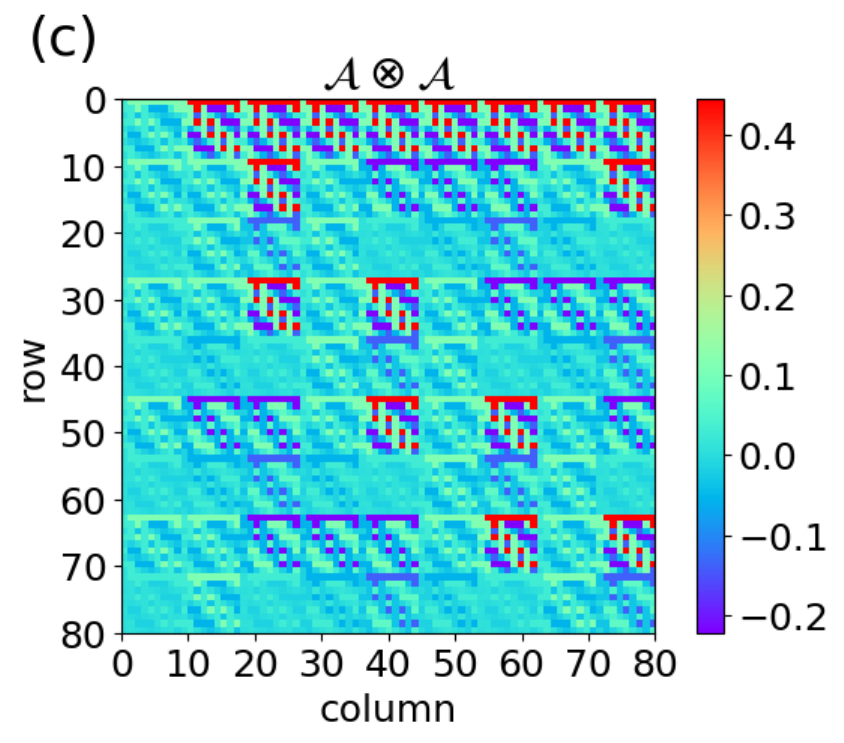}
\caption{The shape and the order of magnitude of the matrices $\mathcal{A}$,$\mathcal{B}$ and $\mathcal{A}\otimes\mathcal{A}$ in (a), (b) and (c) respectively. 
Together they form the matrices \eqref{eq:matrixC} and \eqref{eq:matrixC_1}.\label{fig:matrix}}
\end{figure}

Figures~\ref{fig:matrix}(a) and (b) show the matrices of 
Eqs.~\eqref{eq:matrixC} and~\eqref{eq:matrixC_1} in the case of a single lattice site respectively. 
We remind that for any lattice with some spatial extension, the matrix $\mathcal{C}$ is 
not further reducible, implying that the number of two-qubit gates grows exponentially with $n$, whereas 
for the single time-step case, the matrix $\mathcal{C}_{s=1}$ is encoded within a fixed number of gates, 
for any number of lattice sites, thus making it exponentially more efficient than any classical algorithm. 

We tested with \text{Qiskit} the number of two-qubit gates needed to construct the 
single step circuit. A number of two-qubit gates of the order of $4^7 =16384$ is needed to 
produce both the controlled $U_a$ and $U_b$ of the circuit, yielding a total value 
about $\sim 30,000$ two-qubit gates. 
Although this number is too large for present-day quantum hardware,  it might become
viable in the near or mid-term. 

As is well known, the downside of any single time-step implementation, not just the Carleman LB
algorithm discussed here, is the overhead due to the embedding and readout processes, that need to be 
repeated at  every single timestep,  thereby spoiling the quantum advantage. 
These are important issues that need to be addressed in future work.  

Looking at Fig.~\ref{fig:matrix}(a), we see the specific symmetries of the 
non-local operator $C$ could lead to a lower number 
of gates, and therefore future work should explore the best compiling method, using 
several potential techniques~\cite{shende_synthesis_2006,rakyta_approaching_2022,madden_best_2022}, for example 
tensor network analysis~\cite{felser_efficient_2021,robertson_approximate_2023} to minimize non-local correlations~\cite{succi_quantum_2023}. 
Another potential route is to apply the CL procedure to the fluid equations in their native 
Navier-Stokes form: the Carleman matrix is seemingly more complex due to the cross-correlations 
between fluid density, flow and pressure,  but likely to entail a lesser number of Carleman variables,
with a consequent benefit on the depth of the algorithm. 
Finally,  also the long-known option of special-purpose quantum hardware might be worth
being revisited \cite{yepez_quantum_2002,succi_lattice_1993}.

\section{Conclusions and outlook}\label{sec:VII}

In this work we have developed a Carleman linearization of the
Lattice Boltzmann dynamics for a weakly-compressible fluid for both classical
and quantum computers.
The most promising result is that the relative error between CL--LBM is well within the
``physiological" level of the standard lattice Boltzmann, method at least for moderate
Reynolds numbers up to $\mathcal{O}(10-100)$.  
Although this value does not describe turbulence, it definitely displays sizeable non-linear effects,
leaving hope that turbulent regimes can be attended in the future,  an hypothesis that can only
be tested on quantum computers. 
 
The CL procedure becomes rapidly unfeasible on classical computers, showing that 
in a classical framework, trading nonlinearity for extra-dimensions is a very inconvenient bargain.  
In fact, the exponential increase of Carleman variables with the truncation 
order makes this method substantially useless for relevant applications, as the number 
of variables quickly reaches the limit of exascale supercomputers even on very small grids.

Nonetheless, we stress that the ultimate goal of CL is to implement the embedding of fluid dynamics
onto quantum computers, where the number of qubits scales like the logarithm 
of the number of variables, thus taming the exponential increase of the number of Carleman variables 
via a suitable embedding onto qubits. 
To this purpose, we have proposed and described the explicit form of the quantum 
circuit implementing the CL procedure as applied to the Lattice Boltzmann formulation of fluids. 
To the best of our knowledge, this is the first work  delivering an explicit formulation
and implementation of the quantum algorithm into an actual quantum circuit. 
Specifically,  we have derived the circuit for both the streaming 
and the collision operators and combined them in terms of global unitary gates. 
The latter circuit faces with a formidable depth problem, scaling like $(NQ)^4$,  seemingly unviable 
on any quantum computer, unless dramatic improvements on error correction/mitigation 
procedures are achieved in the coming years.
However, this steep barrier can be tamed by turning to single-step formulations,  featuring
 a fixed circuit depth, regardless of the number of lattice sites. 
Developing an efficient multi-step formulation stands out as a major 
challenge for the Carleman approach to quantum computing of fluids.

Among prospective directions to be explored, we mention the Carleman procedure applied to the
native Navier-Stokes formulations,  concrete applications of the Solovay-Kitaev theorem, 
tensor-network analysis or the use of parametric quantum circuits.
Finally,  special-purpose quantum computers for fluid might also be worth revisiting.  

\begin{acknowledgments}
We acknowledge financial support from National Centre for HPC, Big Data and Quantum Computing (Spoke 10, CN00000013). We also acknowledge the CERN and IBM  Quantum Hub with which the Italian Institute of Technology (IIT) is affiliated.
The authors gratefully acknowledge discussions with M. Maronese, A. Solfanelli, R. Steijl, K. Sreenivasan and W. Itani.
\end{acknowledgments}
The authors have no conflicts to disclose.
The data that support the findings of this study are available from the corresponding author upon reasonable request.
\bibliographystyle{unsrt}
\bibliography{Bibliography}

\end{document}